\documentclass[epj]{svjour}
\usepackage{epsfig}
\usepackage{color}
\newcommand{\eq}[1]{Eq.~\ref{#1}}
\newcommand{\ful}{\mbox{C$_{\scriptsize 60}$}}
\newcommand{\fulp}{\mbox{${\mbox{C}_{\scriptsize 60}}^+$}}
\newcommand{\fuls}{\mbox{\scriptsize C$_{60}$}}
\newcommand{\fulps}{\mbox{\scriptsize C$_{60}^+$}}

\begin{document}
\title{A density functional theory based comparative study of hybrid photoemissions from Cl@$\ful$, Br@$\ful$ and I@$\ful$} 
\titlerunning{A DFT study of hybrid photoemission among Cl@, Br@ and I@$\ful$}
\author{Dakota Shields\inst{1} \and Ruma De\inst{1} \and Esam Ali\inst{1} \and Mohamed E Madjet\inst{2} \and Steven T Manson\inst{3} \and Himadri S. Chakraborty\inst{1} 
}                     % Do not remove
\institute{Department of Natural Sciences, D.L. Hubbard Center for Innovation, Northwest Missouri State University, Maryville, Missouri 64468, USA, \email{himadri@nwmissouri.edu} \and Qatar Environment and Energy Research Institute, Hamad Bin Khalifa University, P.O. Box 34110, Doha, Qatar \and Department of Physics and Astronomy, Georgia State University, Atlanta, Georgia, USA}
\date{Received: date / Revised version: date}
%% The correct dates will be entered by Springer
%
\abstract{Photoionization from atom-$\ful$ hybrid levels in halogen endufullerene molecules, Cl@$\ful$, Br@$\ful$ and I@$\ful$, are calculated using a linear response density functional method. Both the ordinary electron-configuration where the open shell halogen is at the center of $\ful$ and the stable configuration after the atom receives an electron from $\ful$ to form a closed shell anion are considered. Similar ground state hybridization is found for all three systems while, in general, a slight weakening of the effect is noticed after the electron transfer. At lower photon energies, cross sections of the outer hybrid levels attain identical shapes from enhancements driven by the $\ful$ plasmon resonances, while the higher energy emissions remain distinguishable from the differences in atomic responses. These results further show near insensitivity to the choice of a configuration. The inner hybrid cross sections in general exhibit similar overall structures, although differ in details between molecules. However, for these states the results significantly differ before and after the electron transfer -- a feature that can be useful to experimentally determine the real configuration of the molecules $via$ photoelectron spectroscopy.}
\maketitle
\section{Introduction}
\label{sec1}
Spectroscopic research on solid phase and gas phase endofulerenes -- an atom or a smaller molecule taken captive inside a fullerene~\cite{popov2013} -- is important to generate a knowledge repository. This may find fundamental use in prospective applications of these nanosystems which include quantum computations~\cite{harneit2007,ju2011}, organic photovoltaics~\cite{ross2009}, superconductivity~\cite{takeda2006} and biomedical sciences~\cite{melanko2009}. Merged beam techniques were employed at the ALS at Berkeley to probe photoionization properties of atomic endofullerenes experimentally \cite{mueller2008,kilcoyne2010,phaneuf2013}. It may be possible in future to employ photoelectron spectroscopy techniques~\cite{ruedel2002} to access level-selective measurements as well.

Theoretical model studies of the photoresponse of closed-shell atomic endofullerenes are aplenty; some accounts can be found in the review articles Refs.\,\cite{chakraborty2015} and \cite{dolmatove2009}. Studies have regularly predicted hybridization between a varied high-lying orbitals of the atom and the fullerene~\cite{chakraborty2009,madjet2010,maser2012,javani2014a,javani2014b}. The photoionization process of these hybrid orbitals, being rich in dynamical character from admixing spectral signatures of both the atom and the fullerene, are special for spectroscopic studies.

Endofullerenes with open-shell atoms, in contrast, are studied rather scantily. On the other hand, due to the existence of unpaired electrons, there are attractive fundamental interests in such systems. These include long spin relaxation times in N@$\ful$~\cite{morton2007} while enhancement and diminution in hyperfine coupling, respectively, in P@$\ful$~\cite{knapp2011} and exotic muonium@$\ful$~\cite{donzelli1996}. For an atom with one outer-shell vacancy other secondary processes can be induced if energetically accessible. For instance, the transfer of a fullerene electron to fill in the atomic vacancy. This will likely result into a stable electronic configuration due to the formation of a closed shell atomic anion. If this configuration is an excited configuration of the system then it will result into a metastable molecular state. Of course, the realistic ground state of the compound may as well be a mixture of both the configurations, before and after the electron transfer. Therefore, the comparison of the photoionization properties between these two diabatically unique configurations, particularly of the hybrid levels with a focus on understanding the interplay between the two configuration modes, can be uniquely interesting. Using a density functional theory (DFT) framework, a recent study of Cl@$\ful$ has been conducted by us to explore the molecule's hybrid emission behavior~\cite{shields2020}. The natural next step is to study endofullerenes with larger halogen atoms which we report in this paper. Detailed results of single-photoiozation from the atom-fullerene hybrid levels of Cl@$\ful$, Br@$\ful$ and I@$\ful$ are compared and analyzed in detail.

Closed-shell Ar, Kr or Xe being chemically inert, almost certainly locate at the center of the spherical $\ful$. We first treat the barely open-shell Cl, Br and I at the center of $\ful$ within a spherical framework. We call them the {\em ordinary} configurations. A reactive halogen atom is very likely to capture an electron from $\ful$ which will likely bring the compound to a more stable configuration by forming closed-shell Cl$^-$, Br$^-$ or I$^-$. Therefore, we also consider systems of Cl$^-$@$\fulp$, Br$^-$@$\fulp$, I$^-$@$\fulp$ produced by the transfer of a $\ful$ electron to fill in the Cl, Br and I valence shell. There has been experimental evidence, based on laser desorption mass spectroscopy, of $\ful$ with a single Cl$^-$ inside~\cite{zhu1994}. While it is expected that the polarization interaction of the ion can induce some offset in its position from the center of $\ful$, a DFT calculation with Born-Oppenheimer molecular dynamics indicates that this offset is quite small for Cl$^-$ and almost zero for Br$^-$ within neutral $\ful$~\cite{pawar2011}. Earlier studies showed only small effects of the cage polarization except very close to the ionization threshold~\cite{dolmatov2010}. Likewise, a relatively weak effect on the process from a small offset of the atomic location was predicted~\cite{baltenkov2003}. Therefore, we treat Cl$^-$@$\fulp$, Br$^-$@$\fulp$, I$^-$@$\fulp$ assuming spherical geometry as well. We then compare the hybrid photoionization for these stable configurations with those for ordinary configurations. 

\section{A succinct theoretical account}
\label{sec2}
The details of the theoretical schemes are described in Ref.\,\cite{madjet2010} and more recently in Ref.\,\cite{shields2020}. Choosing the photon polarization along the $z$-axis, the photoionization dipole transition cross section in a linear response approximation of time-dependent DFT is given by,
\begin{equation}\label{cross-pi}
\sigma_{n\ell\rightarrow k\ell'} \sim |\langle \psi_{\mathbf{k}\ell'}|z+\delta V|\phi_{n\ell}\rangle|^2.
\end{equation}
Here $\mathbf{k}$ is the momentum of the continuum electron, $z$ is the one-body dipole operator, $\phi_{nl}$ is the single electron bound wavefunction of the target level, and $\psi_{\mathbf{k}l'}$ is the respective outgoing dipole-allowed continuum wavefunction, with $l'=l\pm1$. $\delta V$ represents the complex induced potential that accounts for electron correlations within the linear response of the electrons to the photon field. The computation of $\delta V$ involves determining photon energy dependent induced change in the electron density to be obtained by varying the ground state potential with respect to the ground state electron density as described in Ref.\,~\cite{madjet2008}.

We model the bound and continuum states, and the ground state potential, using the independent particle DFT approximation that utilizes the Leeuwen-Baerends (LB) exchange-correlation functional~\cite{van1994exchange}. This functional involves the gradient of the electron density in the scheme described earlier~\cite{choi2017}. A core of 60 C$^{4+}$ ions for $\ful$ is constructed by smearing the total positive charge over a spherical shell with known molecular radius $R = 6.70$ a.u.\ ($3.54 \AA$)~\cite{ruedel2002} and thickness $\Delta$. The Kohn-Sham equations for the system of 240 $\ful$ electrons (four valence $2s^22p^2$ electrons from each carbon atom), {\em plus} all electrons of the central atom/ion, are then solved self-consistently. The values of $\Delta$ and a pseudo potential used are determined both by requiring charge neutrality and obtaining the experimental value~\cite{devries1992} of the first ionization threshold of $\ful$. $\Delta = 2.46$ a.u.\ ($1.30 \AA$) thus obtained closely agree with the value extracted from measurements~\cite{ruedel2002}. Within the same framework, we also selectively omit either the atom (anion) or $\ful$ ($\fulp$) to obtain the corresponding empty $\ful$ ($\fulp$) and free atomic (anionic) results. For empty fullerenes the model describe two single electron bands in the ground state with one of $\sigma$ (no radial node) and another of $\pi$ (one radial node) character~\cite{choi2017}.

In the density functional model as in the current study, one may adjust the parameters of the functional, or use a different functional, to force accurate ground state properties~\cite{anderson2017}. However, we could not find a unique set of parameter values of our LB functional to work for both free Cl, Br, I and empty $\ful$. Therefore, we have used the set, that is successful for $\ful$, for the endohedral composite-systems as well. The same parametric values were also used to calculate results of atomic anions and $\fulp$ included in the discussion.  But these values, for instance, overestimate Cl ionization potential of NIST database~\cite{kramida2018} by about 7\%. They further produce an electron affinity of 4.15 eV as opposed to the measured value of 3.6 eV~\cite{berznish1995} for Cl. The same level of small inaccuracies are also noted for Br and I, that can be seen in Table I. But these small inaccuracies should not take away much from the main results of this study which explores the dominant effects of $\ful$. 
\begin{table}
\caption{Ionization potential (IP) and electron affinity (EA) calculated. The values in the parenthesis are NIST data.}
\label{tab2}       % Give a unique label
\begin{tabular}{c|cc}
\hline\noalign{\smallskip}
              & IP (eV)   & EA (eV) \\
\noalign{\smallskip}\hline\noalign{\smallskip}
Cl   & 13.9 (13.0; Ref.\,\cite{kramida2018})  & 4.15 (3.60; Ref.\,\cite{berznish1995})   \\
Br   & 13.0 (11.8; Ref.\,\cite{lide1992})  & 4.10 (3.36; Ref.\,\cite{blondel1992})   \\
I    & 11.8 (10.5; Ref.\,\cite{lide1992})  & 4.06 (3.06; Ref.\,\cite{Pelaez2009})   \\
\noalign{\smallskip}\hline
\end{tabular}
% Or use
%\vspace*{1cm}  % with the correct table height
\end{table}
\begin{figure}
\vskip 0.0 cm
\centerline{\psfig{figure=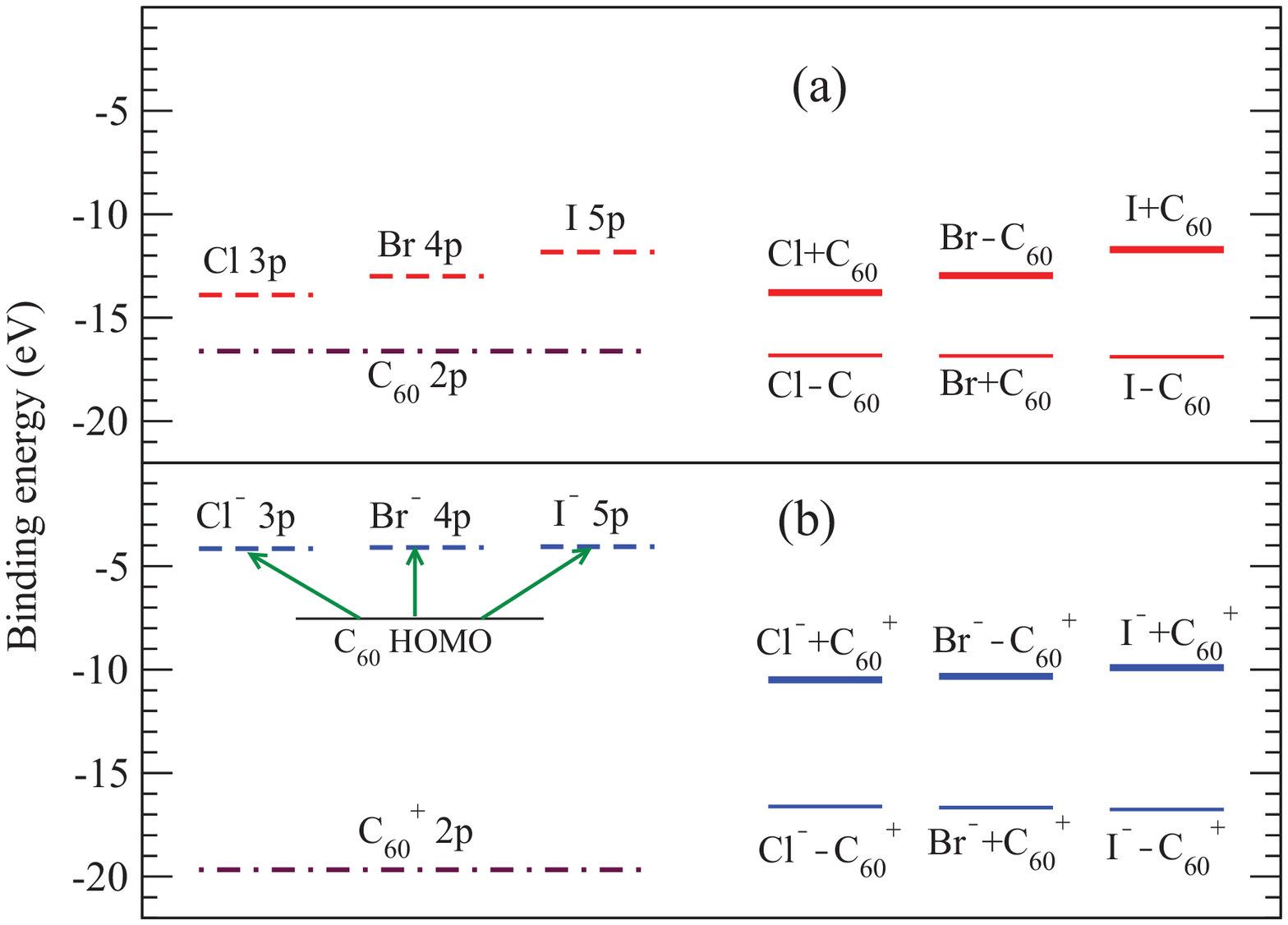,height=8.5cm,width=9.5cm,angle=0}}
%\resizebox{0.75\textwidth}{!}{%
%  \includegraphics{Fig1-prodensity.eps}
%\vspace{5cm}       % Give the correct figure height in cm
\caption{(Color online) Valence $np$ level binding energies of free halogen atoms (a) and anions (b), and that of the $2p$ levels of empty $\ful$ (a) and $\fulp$ (b) are drawn on the left side. The corresponding symmetric and antisymmetric hybrid levels of the compound X@$\ful$ (a) and X$^-$@$\fulp$ (b) are presented on the right side. Electron transition energies from $\ful$ HOMO to the halogen's $np$ hole are indicated on panel (b).}
\label{fig1}       % Give a unique label
\end{figure}

\section{Results and discussion}
\label{sec3}
\subsection{Ground State Hybridization}

One useful way to describe hybridization in electronic states of a multi-member system is the assessment of both the level energy separation and the wavefunction overlap between states of the participating members. Both the decrease of the former and the increase of the latter favor hybridization. To that end, Fig.\,1 presents detailed energy information as obtained from the current calculations. Note that from the orthogonality of wavefunctions in spherical systems, only states with same angular momentum can hybridize. The left side of Fig.\,1(a) presents the empty $\ful$ $2p$ level energy and $np$ energies of X = Cl, Br and I. We use Coulomb notation for the atomic levels and harmonic oscillator notation for the $\ful$ levels. Note that the $np$ binding energy systematically decreases going from Cl to I thereby increasing the energy separation from $2p$ $\ful$. It is then expected that $2p$ of the $\fulp$ cation will get more bound while the energy levels of the X$^-$ anions will become less bound compared to their neutral counterparts. This is seen on the the left side of Fig.\,1(b), where it is evident that the loosening of $np$ levels going from Cl$^-$ to I$^-$ is almost negligible. Furthermore, the $np$ energies of anions are practically the electron affinity values of the neutrals (Table I). Energies of the resultant hybrid states, for both X@$\ful$ and X$^-$@$\fulp$ configurations of composite systems, are given on the right sides of both panels where + and - signs represent symmetric (bonding) and antisymmetric (antibonding) hybrids respectively. It is interesting to note that for both configurations the higher binding energy (inner) hybrid levels have practically the same binding energy. On the other hand, for each configuration the outer hybrid levels systematically move higher from Cl to I, while the X@$\ful$ configuration shows an weaker change with an overall lower binding. 
\begin{figure}
\vskip 0.7 cm
\centerline{\psfig{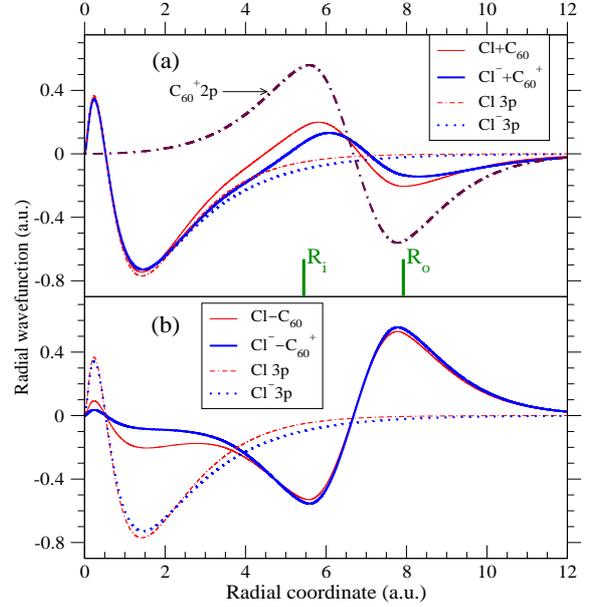}}
%\resizebox{0.75\textwidth}{!}{%
%  \includegraphics{Fig1-prodensity.eps}
%\vspace{5cm}       % Give the correct figure height in cm
\caption{(Color online) Radial symmetric (a) and antisymmetric (b) hybrid wavefunctions of Cl@$\ful$ \textit{versus} Cl$^-$@$\fulp$. For both the molecules the $3p$ level of the free atom Cl and anion Cl$^-$ hybridizes with the $2p$ level of, respectively, empty $\ful$ and $\fulp$. Corresponding wavefunctions of free Cl and Cl$^-$ are displayed, while that of only $\fulp$ is shown on panel (a). The inner ($R_i$) and outer ($R_o$) radii of the $\ful$ shell are shown on panel (a).}
\label{fig2}       % Give a unique label
\end{figure}
\begin{figure}
\vskip 0.7 cm
\centerline{\psfig{figure=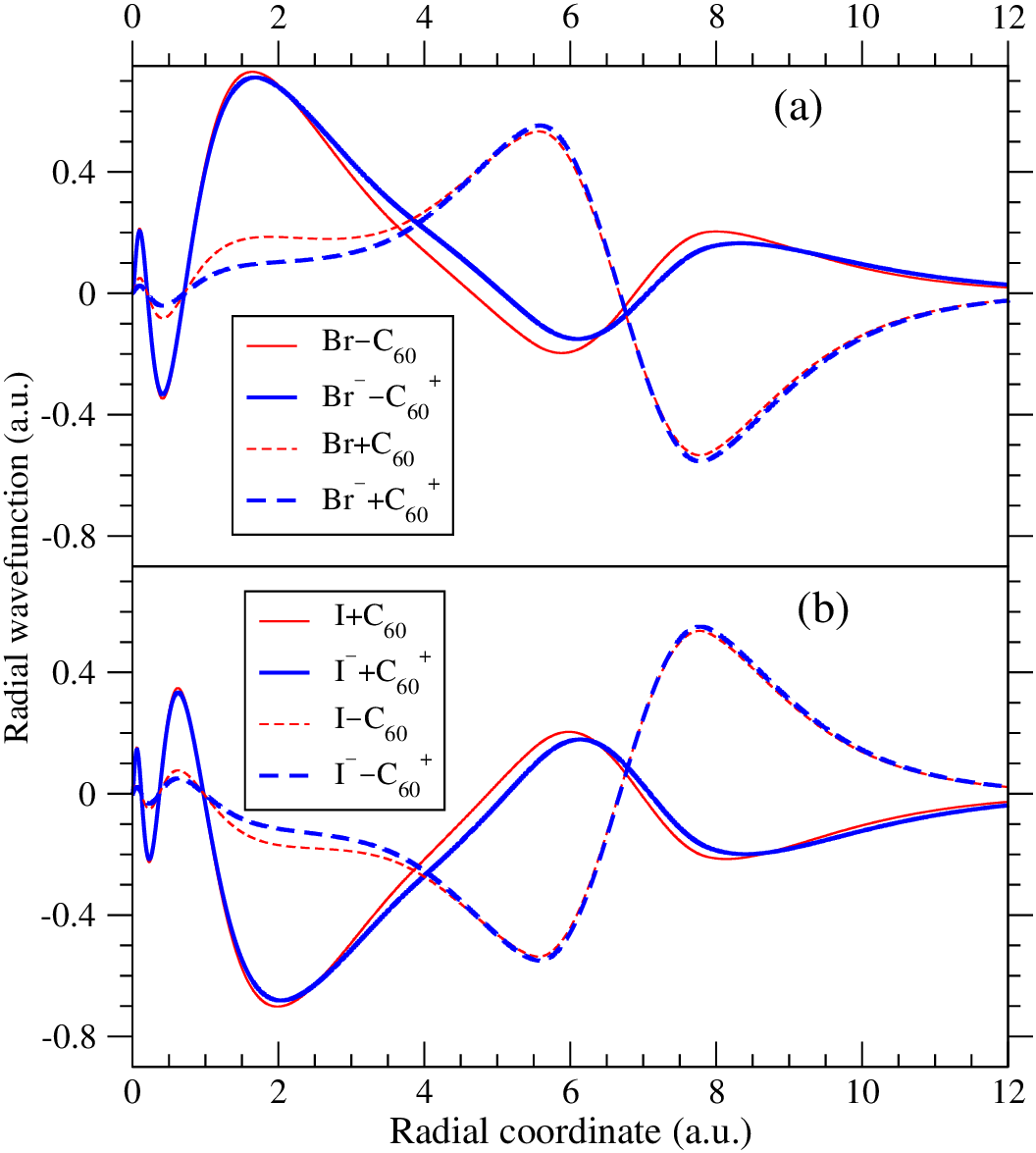,height=8.0cm,width=7.5cm,angle=0}}
%\resizebox{0.75\textwidth}{!}{%
%  \includegraphics{Fig1-prodensity.eps}
%\vspace{5cm}       % Give the correct figure height in cm
\caption{(Color online) Radial hybrid wavefunctuions of Br@$\ful$ \textit{versus} Br$^-$@$\fulp$ (a) and I@$\ful$ \textit{versus} I$^-$@$\fulp$ (b) systems.}
\label{fig2}       % Give a unique label
\end{figure}

As noted above, the configuration of simply placing a neutral halogen at the center of a neutral $\ful$ is ordinary. For $\fulp$, due to large cloud of 240 delocalized electrons, the ground state structure is insensitive to the location of the hole among the molecular levels. Likewise, for X$^-$@$\fulp$ the level energies and wavefunctions of pure $\ful$ and hybrid states are found to be independent of at which $\ful$ orbital the hole is situated. Let us consider an electron transition from the highest occupied (HOMO) level of $\ful$ of binding energy -7.52 eV to the outer $np$ of X of electron affinity (see Table I) which is the $np$ binding energy of X$^-$ as noted above. These transitions are indicated in Fig.\,1(b). The resulting configuration with the HOMO vacancy will have the minimum total energy and, therefore, will be the ground state configuration for X$^-$@$\fulp$. This does not even take into account the extra binding associated with the electrostatic Coulomb attraction between X$^-$ and $\fulp$. Despite the electron affinity of Cl being less than the ionization potential of $\ful$ (see above), this extra binding is what that will enable the ionic compound to bind. Therefore, this ground state configuration of Cl$^-$@$\fulp$ should be stable and abundantly formed. Simultaneously, if this energy is still higher than the total energy of X@$\ful$, then the configuration will be metastable and of significant spectroscopic interest. One way to probe the physical situation is to conduct photoelectron spectroscopic measurements that may test the state selective ionization results described in the following subsection.

The Radial wavefunctions corresponding to two configurations of the Cl system are presented in Fig.\,2 with panel (a) and (b) for, respectively, symmetric and antisymmetric hybrids. In both panels the $3p$ wavefunctions of Cl and Cl$^-$ and that of $2p$ for $\fulp$ are shown. Note that $2p$ is $\pi$-type fullerene state. For the hybrid states, a somewhat weakening of hybridization is noted in going from Cl@$\ful$ to Cl$^-$@$\fulp$. However, the extent of hybridization still achieved for Cl$^-$@$\fulp$ in spite of very large separation [Fig.\,1(b)] between free Cl$^-$ and empty $\fulp$ states is surprising at a first look. However, some outward radial shift (Fig.\, 2) of $3p$ wavefunction, Cl$^-$ \textit{versus} Cl, and a small inward shift of $2p$ of $\fulp$ compared to $\ful$ (not shown), likely advantaged the mixing. The situation is qualitatively similar for Br and I systems whose hybrid wavefunctions are presented in Fig.\,3(a) and 3(b) respectively. Note, however, that due to the two-node character of the Br (Br$^-$) $4p$ wavefunction, its symmetric combination with one-node $\ful$ ($\fulp$) $2p$ will have one less node than the antisymmetric combination to become the inner hybrid, while for Cl $3p$ and I $5p$ having one and three nodes reverse this order. Finally, the hybrid states can be symbolically expressed as,
\begin{equation}\label{bound-hyb}
|\mbox{X}\pm\mbox{C}_{60}\rangle = |\phi_\pm\rangle = \eta_\pm|\phi_{np \mbox{\scriptsize X}}\rangle \pm \eta_\mp|\phi_{2p \fuls}\rangle
\end{equation}
for X@$\ful$, and the same for X$^-$@$\fulp$ with X and $\ful$ replaced by X$^-$ and $\fulp$. Here $\eta_+ = \sqrt{\alpha}$ and $\eta_- = \sqrt{1-\alpha}$ where $\alpha$ is the mixing parameter that renders the hybrid states orthonormal.  

\subsection{Hybrid Photoionization}
\begin{figure}
\vskip 0.7 cm
\centerline{\psfig{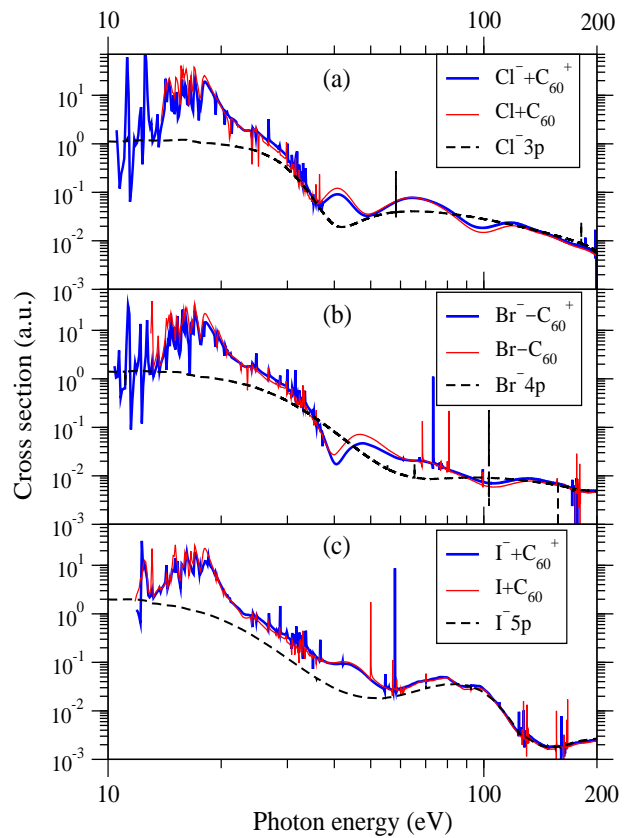}}
%\resizebox{0.75\textwidth}{!}{%
%  \includegraphics{Fig1-prodensity.eps}
%\vspace{5cm}       % Give the correct figure height in cm
\caption{(Color online) Photoionization cross sections of the outer hybrid electron in Cl@$\ful$ \textit{versus} Cl$^-$@$\fulp$ (a), Br@$\ful$ \textit{versus} Br$^-$@$\fulp$ (b), and I@$\ful$ \textit{versus} I$^-$@$\fulp$ (c). The cross section of the outer $np$ ionization of the corresponding free anion is also presented on each panel for comparisons.}
\label{fig3}       % Give a unique label
\end{figure}
\begin{figure}
\vskip 0.7 cm
\centerline{\psfig{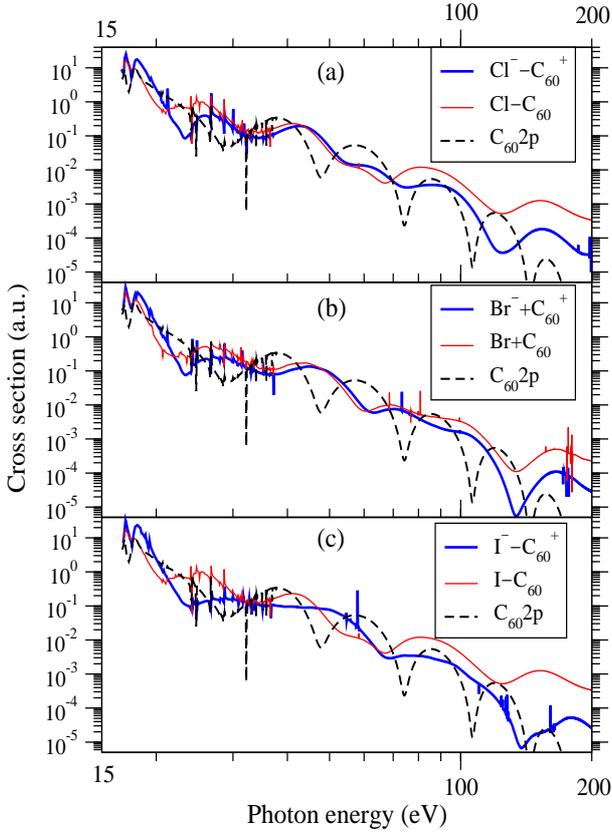}}
%\resizebox{0.75\textwidth}{!}{%
%  \includegraphics{Fig1-prodensity.eps}
%\vspace{5cm}       % Give the correct figure height in cm
\caption{(Color online) Same as Fig.\,4 but for the photoionization of the inner hybrid levels.}
\label{fig4}       % Give a unique label
\end{figure}
%.

\subsubsection{Emissions from outer hybrid level}

Cross sections calculated in linear response time-dependent DFT for the outer hybrid levels for both configurations are presented in in Fig.\,4 with panels (a), (b) and (c), respectively, for Cl, Br and I composites. Several narrow resonances are due to the many autoionizing resonance channels, driven by Auger and inter-Coulombic decay (ICD) process~\cite{javani2014b}, that exist in the endofullerenes due to levels in the fullerene molecule that are not there in free atoms. In any case, comparing the curves with the $np$ cross sections of free X (not shown) and X$^-$, which are practically same over these energies, indicates plasmon driven enhancements at lower photon energies up to 30 eV~\cite{chakraborty2009,javani2014a,madjet2007,javani2014}. In the framework of interchannel coupling due to Fano \cite{fano1961}, the correlation-modified matrix element of the photoionization of X$\pm\ful$ can be written perturbatively as \cite{javani2014a,shields2020},
\begin{eqnarray}\label{gen-mat-element}
{\cal M}_\pm (E) &=& {\cal D}_\pm (E)\nonumber \\
                 &\!+\!\!\!& \!\!\displaystyle\sum_{n\ell}\!\!\int \!\!\!dE' \!\frac{\langle\psi_{n\ell}(E')|\frac{1}{|{\bf r}_{\pm}-{\bf r}_{n\ell}|}
|\psi_{\pm}(E)\rangle}{E-E'} {\cal D}_{n\ell} (E')
\end{eqnarray}
in which the single electron (uncorrelated) matrix element, that is the matrix element without $\delta V$ in \eq{cross-pi}, is
\begin{equation}\label{se-mat-element}
{\cal D}_\pm (E) = \langle ks(d)|z|\phi_\pm\rangle
\end{equation}
and $|\psi_{nl}\rangle$ in the interchannel coupling integral is the (continuum) wavefunction of the $n\ell\rightarrow k\ell'$ channel. Taking the hybridization into account, the channel wavefunctions in \eq{gen-mat-element} become
\begin{equation}\label{channel-hyb}
|\psi_\pm\rangle = \eta_\pm|\psi_{np \mbox{\scriptsize X}}\rangle \pm \eta_\mp|\psi_{2p \fuls}\rangle
\end{equation}
  
Substituting Eqs.\ (\ref{bound-hyb}) and respective Eqs.\ (\ref{channel-hyb}) in \eq{gen-mat-element}, and noting that the overlap between a pure X (X$^-$) and a pure $\ful$ ($\fulp$) bound state is negligible, we separate the atomic and fullerene contributions to the integral to get the full (correlated) matrix element for X$\pm\ful$ and X$^-\pm\fulp$ levels as,
\begin{equation}\label{gen-mat-element}
{\cal M}_\pm (E) = \eta_\pm{\cal M}_{np \mbox{\scriptsize X(X$^-$)}} (E) \pm \eta_\mp{\cal M}_{2p \fuls(\fulps)} (E)
\end{equation}
where the first and second terms, respectively, on the right hand side describes interchannel coupling effects of atomic and fullerene ionization channels. 

A large number of fullerene channels, which are very strong due to the photoionization of plasmon resonances, exist at lower energies. Through the second term in \eq{gen-mat-element}, these channels couple with the $np$ emissions of comparable strengths from X and X$^-$ (Fig.\,4). This explains the almost similar cross sections in broad shapes and magnitudes over these energies making the results insensitive to the choice of the system or the configuration. The broad shoulder structures above 20 eV is likely the effect of the higher energy plasmon resonance~\cite{scully2005}. Note, however, that the levels of X$^-\pm\fulp$ open at lower photon energies since they bind weakly (Fig.\,1). Further note that the opposite symmetry of Cl and I \textit{versus} Br endofullerenes, pointed out above, bears little effect on the results.

As the plasmonic effect weakens with increasing energy, the cross sections (Fig.\,4) largely follow their free atom (ion) curves, since the first term of \eq{gen-mat-element} begins to dominate. The results also show a series of oscillations. These oscillations are a consequence of a well-known multipath interference mechanism~\cite{cdm2000} due to the cavity structure of $\ful$ which was modeled earlier in detail in Ref.\ \cite{mccune2009}. Since the free Cl, Br and I results are distinctly different at higher energies primarily due to the occurrence of various Cooper minima, the outer hybrid results also maintain differences. However, owing to the cross sections of X \textit{versus} X$^-$ being so close, results are seen to be essentially independent to the choice of the configuration.

\subsubsection{Emissions from inner hybrid level}

Fig.\, 5 delineates results of the emissions from the inner hybrid level. Since these channel opens above 15 eV, the effect of the giant plasmon only exists over a small energy range above the ionization threshold, although the effects of the higher energy plasmon linger on slightly further in energy. We, however, note significant differences between the choice of configuration at these lower energies which, however, tend to match at the intermediate energy range. Due to the stronger fullerene charter of these levels (see Figs.\, 2 and 3), causing their lower average-magnitudes, the oscillatory features are found to be more intense compared to the outer hybrid cases. Note that the cross sections for the two choices of the configurations begin to fall off significantly again past 70 eV with the results for X$^-$@$\fulp$ staying lower. 

In Fig.\,5 we further compare the results with the cross section of the $2p$ level of $\ful$ which shows sharper oscillations with very rapid fall-off due to the absence of any atomic type steady emission~\cite{shields2020,mccune2008}. This comparison exhibits a stronger non-oscillatory background strength for inner hybrid emissions that largely weakens the sharpness of the oscillations. The reason for this is the contributions to the amplitude from the atomic region owing to the structures that exist there in the wavefunction due to the hybridization, as can be seen in Figs.\, 2 and 3. This can be understood in the dipole acceleration gauge formalism of the ionization amplitude which involves the potential gradient~\cite{potter2010}. The derivative of the Coulomb-type potential at the center is large, since $\frac{d}{dr}(-1/r) = \frac{1}{r^2}$. As a result, even a small probability density at the center (due to hybridization) can create significant contributions to the matrix element. This will be more prominent at higher energies. Therefore, electron probability densities, however small, at the central region receive strong recoil force from the Coulombic potential ridge to augment the matrix element.

\section{Conclusion}
\label{sec4}
Using a DFT methodology with the LB exchange correlation functional, ground state atom-$\ful$ single electron hybrid levels of $p$ angular momentum character are predicted for Cl, Br and I centered endofullerene molecules. Single electron dipole photoionization of these systems are investigated within the framework of linear response time-dependent DFT. While the degree of hybridization in all three molecules is found similar, the effect somewhat reduces after a $\ful$ electron transfers to occupy the halogen vacancy producing likely a more stable configuration. For the outer hybrid states the ionization response over the plasmonic range of the spectra is almost indistinguishable among all systems, including before and after the electron transitions, barring the detailed structures of narrow autoionizing resonances. At higher energies likewise, while little sensitivity arises from the electron's relocalization, the confined halogen's character dominates the ionization of the outer hybrid. The ionization of the inner hybrid levels at lower energies, in contrast, modifies substantially upon the electron transfer. This difference diminishes considerably at intermediate energies to again become important as the energy further increases. Ionization for these hybrids, even though of dominant $\ful$ character, however, draws some extra strength from the atomic zone. The details of these differences and similarities, along with the delineation of the actual electron configuration of the molecules, are excellent candidate for study $via$ photoelectron spectroscopic experiments.

The $\ful$ ion core is smeared over a jellium sphere in our model that freezes bond vibrations. A moot question therefore remains. Can the oven temperature of about 800 K to produce fullerene vapor wash out the broader structures discussed in this paper? Sample temperature can affect the situation in two ways: (i) coupling of the electronic motion with the temperature-induced vibration modes of the ion core~\cite{bertsch1989} and (ii) fluctuation of the cluster shape around the shape at absolute zero~\cite{pacheco1989}. However, as shown in Ref.\ \cite{madjet2008}, it required a convolution of the theoretical results to add a width less than 1 eV to compare with photoionization measurements of gas phase $\ful$. This width is rather miniscule in comparison with energy resolution of about 5-10 eV required to measure broad structures in Figs.\,4 and 5. Therefore, thermal vibrations, while will likely smooth the autoionizing spikes, will not qualitatively alter the key results presented.

\begin{acknowledgement}
{\bf ACKNOWLEDGEMENT:} The research is supported by the US National Science Foundation Grant No.\ PHY-1806206 (HSC) and the US Department of Energy, Office of Science, Basic Energy Sciences, under award DE-FG02-03ER15428 (STM).
\end{acknowledgement}
{\bf Author Contributions:} STM and HSC conceived the problem; MEM and HSC designed and implemented the research; DS, RD and EA contributed to the computations, while all authors contributed to the analysis of the results; RD, STM and HSC primarily worked to write the manuscript.
%
% BibTeX users please use
% \bibliographystyle{}
% \bibliography{}

\end{document}